\documentclass[12pt,preprint]{aastex}

\shorttitle{IRS spectroscopy of k+a galaxies}
\shortauthors{Roseboom et al.}

\begin{document}

\title{{\it Spitzer} IRS observations of k+a galaxies: A link between PAH emission properties and AGN feedback?}

\author{I.G. Roseboom\altaffilmark{1}, S. Oliver\altaffilmark{1}, D. Farrah \altaffilmark{1}}

\altaffiltext{1}{Dept of Physics \& Astronomy, University of Sussex, Falmer, BN1 9QH, United Kingdom}

\begin{abstract}
We have performed IRS low resolution 5--12 micron spectroscopy on a sample of galaxies selected to be at 3 distinct post-starburst evolutionary stages based on their optical spectral indices. The resulting IRS spectra show distinctive PAH emission line structures at 6.2, 7.7, 8.6 and 11.3$\mu$m and little silicate absorption, indicative of ongoing star formation. However the PAH inter-line ratios, in particular the 11.3/6.2$\mu$m and 7.7/6.2$\mu$m ratio, show large variations. These variations are found to correlate with both time since the most recent starburst and AGN activity. We speculate that the evolution observed in these PAH ratios is related to an increase in AGN activity with time since star burst.
\end{abstract}


\section{Introduction}
The nature of the connection between starbursts and AGN has been long debated. While the effectiveness of AGN feedback in quenching star formation has been heralded as a key component of galaxy formation by semi-analytic models (e.g. \citealt{Croton2006,Delucia2007}), it is only very recently that observational evidence of AGN activity co-incident with post-starburst signatures has been found \citep{Schawinski2007,Bundy2008,Kaviraj2008,Schawinski2008}. However these discoveries form far from a consensus, with some suggestions that AGN feedback alone may not be sufficient for the truncation of star formation in these galaxies \citep{Bundy2008,Kaviraj2008}.

The mid-IR contains unique diagnostics of the star formation, AGN activity, and dust obscuration levels of galaxies. Rest-frame $5-12\mu$m observations can decipher the relative strength of these attributes via investigation of the PAH emission properties, continuum slope, and silicate absorption strength, respectively.

Here we utilise {\it Spitzer} InfraRed Spectrograph (IRS) observations of a sample of galaxies which appear at different stages of post-starburst evolution as determined by their optical spectra. In particular we use the H$\delta$, [OII]3272\AA\/ and 4000$\AA\/$ break spectral features to identify k+a galaxies at 3 different epochs. k+a galaxies are typically identified via their optical spectrum, with strong absorption in the Balmer series lines, but no significant emission in the [OII]3727 \AA\/ line or other emission lines associated with star formation (Dressler \& Gunn 1983; Couch and Sharples 1987; among others). This kind of optical spectrum is indicative of a galaxy which has undergone a very recent burst of star formation as the young, short-lived ($<1-2$ Gyr) A and B stars are still observed, but no ongoing, or very low level, star formation. Spectral modelling of k+a galaxies typically suggests a star-burst in which 1-10\% of the galactic stellar mass is in the form of new stars (i.e. Balogh et al 1999\nocite{balogh99}).

\section{Data}
We selected 12 galaxies in three categories from the overlap between the {\it Spitzer} SWIRE survey and the SDSS spectroscopic dataset for observations with IRS. The targets were selected via their [OII]3727\AA, H$\delta$ (4101\AA), and D4000 indices to be at distinct post-starburst stages. The H$\delta$ and D4000 indices are well established as good indicators of stellar age (i.e. \citealt{balogh99,Kauffmann2003}), while the [OII]3727\AA\/ is a well known indicator of current star formation. Measurements of the indices were performed as per Roseboom et al. (2006)\nocite{Roseboom2006}. To determine our selection criteria we turn to the stellar population synthesis models of \citet{Bruz03}. Models which assume an exponentially declining starburst ($\tau=100$ Myr) on top of a passively evolving single stellar population of 11 Gyr were generated using the \citet{Bruz03} code. Burst strengths of 1,5 \& 10$\%$ were considered. From these models we determine 3 sets of H$\delta$ \& D4000 criteria which isolate galaxies with roughly 0.03, 0.3 and 1 Gyr of post-starburst evolution. In the case of the later two selections an additional [OII] selection cut is imposed to filter out galaxies with significant ongoing star formation. The specifics of this selection are;\\

\vspace{0.5cm}
\begin{tabular}{llll}
Line & Young k+a & k+a & Old k+a\\
\hline
[OII]3727\AA (\AA)$^{*}$& -- & $<6$ & $<6$\\
H$\delta$ (\AA)& $>4$ & $>5$ & $>2$\\
D4000 & $1-1.2$ & $1.3-1.5$ & $1.6-1.8$\\
Time since burst (Myr) & 25 & 300 & 1000\\
\end{tabular}
\\
{$^{*}$In emission}
\vspace{0.5cm}

Observations were performed with the short-low module in staring mode, two nod positions per spectral order with an offset of 20'' were observed with ramp times between 400-600s. The data were reduced using a combination of the SMART (Hidgon et al. 2004) and SPICE software environments. First individual ramps of the same order and nod were combined in SMART utilising a median filtering algorithm. Sky subtraction is performed via subtraction of different order observations at the same nod position. Once combined and corrected for background/sky contributions the spectra are extracted using the SPICE pipeline assuming these objects are point sources for {\it Spitzer} at these wavelengths. Finally the two nods are combined in SMART to produce the final spectrum.

11 of the 12 IRS spectra are found to have acceptable S/N spectra ($> 5$)  however one of the IRS observations appears to have featureless, flat spectrum, suggesting that the exposure times were not sufficient to detect the mid-IR features, or the IRS slit was misplaced in relation to the mid-IR emission. This failed observations is omitted in the following discussion.
\section{Results}
\begin{figure}
\centering
\includegraphics[scale=0.5]{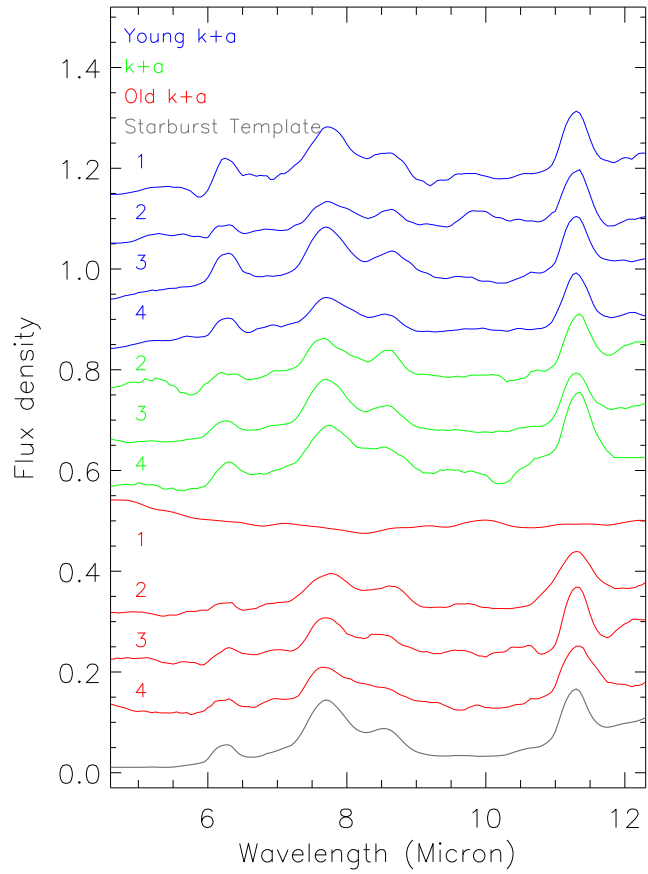}
\caption{IRS SL spectra for the three sub-samples under investigation here. For comparison the starburst template of Brandl et al. (2006) is also shown. Spectra are normalised to arbitrary flux at 11.3$\mu$m for the sake of clarity.}
\label{fig:irs_spec}
\end{figure}

Figure \ref{fig:irs_spec} shows the IRS spectra for the 11 sources discussed here. For comparison the starburst template of Brandl et al. (2006) is also shown.  All the spectra, with the exception of Oka 1, show clear PAH emission features at 6.2, 7.7 and 11.3$\mu$m. Oka 1 is found to have a completely stellar dominated thermal spectrum, similar to early-type galaxies observed with IRS (e.g. NGC 4552; Smith et al. 2007). Negligible silicate 9.7$\mu$m absorption is seen in all cases. %A faint absorption feature at $\sim$5.8$\mu$m is seen a number of the spectra (yka1,ka2,ka4) which could from absorption from ice. However in all cases the feature is only seen in one of the slit positions, and so it is likely an artifact of the reduction as opposed to a real feature.

\begin{figure}
\centering
\includegraphics[scale=0.5]{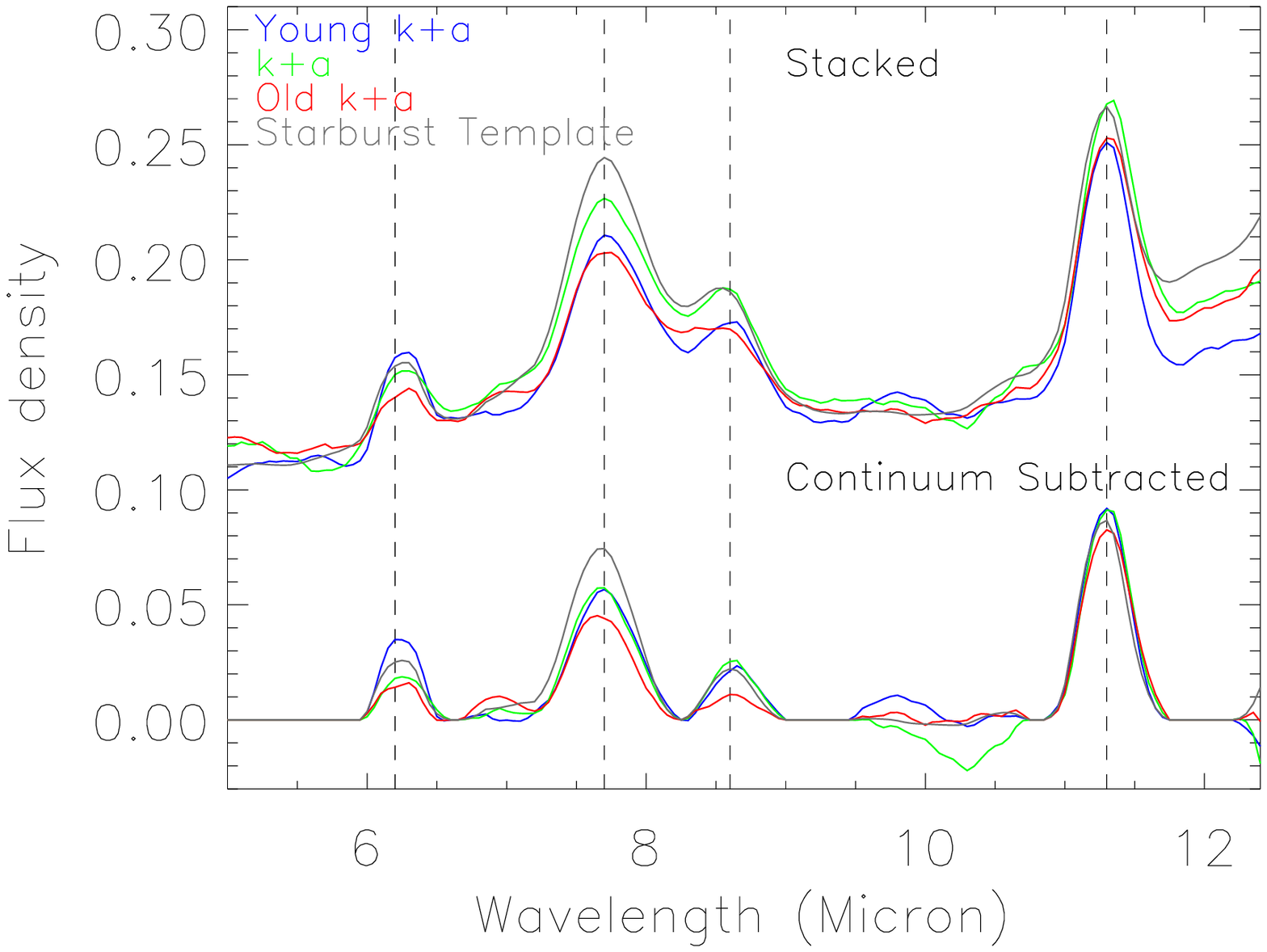}
\caption{Stacked (top) and continuum subtracted (bottom) IRS SL spectra for the three sub-samples under investigation here. Oka 1 is omitted from the old k+a stack as it is clearly AGN dominated. For comparison the starburst template of Brandl et al. (2006) is also shown. Spectra are normalised to arbitrary flux in the 11.3$\mu$m line for the sake of clarity.}
\label{fig:irs_stack}
\end{figure}

Overall the spectra are typical of normal starburst galaxies (i.e. \citealt{Brandl2006}). However one unusual feature of these spectra is the dominance of the 11.3$\mu$m PAH feature over the 6.2$\mu$m PAH feature in the k+a and old k+a types. To emphasis this point in Figure \ref{fig:irs_stack} we show the spectra stacked by type, and also the stacked \& continuum subtracted IRS spectra. 
%This is particularly unusual as for typical ULIRGs with PAH dominate spectra the 6.2$\mu$m feature is the strongest (?), with little variation in the line ratios across large samples (?). This aspect of the spectra will be discussed further in Section \ref{sec:discussion}. 

For each object key optical emission line fluxes (H$\beta$, [OIII]5007, H$\alpha$ and [NII]) are determined by fitting the residual emission after removal of best fit stellar population models using the GANDALF IDL code \citep{Sarzi2006}. 

Equivalent widths (EWs) and fluxes for the PAH features in the IRS spectra are measured via use of the IDL PAHFIT tool\citep{Smith2007}. As the mid-IR coverage is limited to 5-12$\mu$m the determination of the continuum around 7-12$\mu$m is troublesome due to the degeneracy between silicate absorption and PAH emission. However as our spectra appear to have no signs of strong silicate absorption around 9.7$\mu$m we constrain the silicate absorption be be zero by running PAHFIT with the {\tt /NO\_EXTINCTION} flag set.

Table \ref{tab:optprop} lists the optical properties, while Table \ref{tab:irprop} lists the mid-IR line fluxes and equivalent widths.

To identify correlations between these observed properties and the post-starburst evolution we try to fit a two component stellar population model to each galaxies spectrum. The models are similar to those used to select the sample; a young stellar population superimposed on a homogeneously old single stellar population. The young component is taken to be a single burst with a exponentially declining star formation rate of $\tau=100$ Myr. The old component is taken to be a homogeneously 11 Gyr old single stellar population. In both cases models are provided by the code of Maraston (2005). To investigate both the time since starburst and the burst fraction a total of 24 steps in burst time from 10 Myr to 5 Gyr, and 16 burst fractions from 0.01 to 0.15 are considered. The affect of dust is included according to the model of Calzetti et al. (2000) \nocite{Calzetti2000} by allowing E(B-V) to vary between 0-0.2 (A$_v\approx$0-1). Finally three steps in metallicity are considered; $Z=0.5 Z_{\sun}$ , $Z=Z_{\sun}$ and $Z=2Z_{sun}$. In total 5760 models are considered. The best fit solution is found via $\chi^2$ fitting to the standard 25 Lick indices as well as broadband fluxes from; Galex NUV \& FUV, SDSS $u,g,r,i,z$ and {\it Spitzer} 3.6$\mu$m \& 4.5$\mu$m. Where an object is undetected the upper limits are incorporated. The burst time and burst fraction are taken to be the best-fit values after marginalising over the other model parameters. 1$\sigma$ errors are taken to be region surrounding the maximum containing the 68\% of the probability distribution. These quantities are quoted in Table \ref{tab:optprop}.

\begin{deluxetable}{llllllllll}
\tabletypesize{\tiny}
\tablewidth{0pc}

\tablecaption{Measured optical properties of sample}

\tablehead{
\colhead{ID}& \colhead{Name}           & \colhead{$z$}      &
\colhead{Burst Time (Gyr)}          & \colhead{Burst strength} & \colhead{H$\beta$\tablenotemark{a}} &\colhead{ [OIII]5007} & \colhead{H$\alpha$} &\colhead{ [NII]}  & \colhead{Type\tablenotemark{b}}
}

\startdata
Young k+a\\
\hline
Yka1 & J103254.44+581638.5 &  0.08 & $0.27^{+0.08}_{0.07}$& 0.3$^{+0.33}_{-0.28}$ & 106.5 & 104.6 & 291.84 & 54.29 & SF \\
Yka2 & J104826.85+564026.8 &  0.07 & 0.13$^{0.03}_{-0.03}$& $0.5^{+0.1}_{-0.3}$ & 141.95 & 146.66 & 450.88 & 92.33 & SF\\
Yka3 & J105326.83+580956.5 &  0.13 & 0.27$^{+0.08}_{-0.07}$& $0.25^{+0.25}_{-0.16}$ & 229.16 & 156.57 & 770.26 & 234.36 & SF\\
Yka4 & J163332.66+404827.0 &  0.08 & $0.16^{+0.04}_{-0.03}$& $0.25^{+0.25}_{-0.16}$ & 187.53 & 350.97 & 657.17 & 102.81 & SF\\
\hline
k+a\\
\hline
%J104220.14+564855.7 &  0.091 \\
Ka2 & J104953.31+564551.3 &  0.182  & $0.74^{+0.21}_{-0.18}$& $0.15^{+0.08}_{-0.03}$ & 2.56 & 23.98 & 17.16 & 44.74 & AGN\\
Ka3 & J110214.89+574630.6 &  0.14 & 0.35$^{+0.1}_{-0.08}$& $0.05^{+0.15}_{-0.04}$ & 358.86 & 102.05 & 1015.24 & 430.44 & HYBRID\\
Ka4 & J104319.20+563151.0 &  0.21 & 0.58$^{+0.16}_{-0.13}$& $0.08^{+0.26}_{-0.05}$ & 6.83 & 23.99 & 197.1 & 135.08 & AGN\\
\hline
Old k+a\\
\hline
Oka1 & J163252.97+402056.1 &  0.07 & 0.95$^{+0.3}_{-0.2}$ & 0.02$^{+0.16}_{-0.04}$ & 19.07 & 22.31 & 15.69 & 30.13 & AGN \\
Oka2 & J104314.82+570941.2 &  0.12 & 3.4$^{+1.0}_{0.7}$& $0.01^{+0.04}_{-0.01}$ & 22.75 & 15.98 & 87.8 & 48.22 & HYBRID\\
Oka3 & J104650.43+575923.9 &  0.18 & $0.74^{+0.2}_{0.18}$& $0.1^{+0.07}_{-0.08}$ & 29.5 & 39.05 & 140.06 & 85.88 & HYBRID\\
Oka4 & J104834.84+583311.3 & 0.13  & $1.2^{+0.3}_{0.3}$& $0.3^{+0.3}_{0.28}$ & 158.3 & 49.25 & 519.12 & 249.84 & HYBRID\\
\enddata

\tablenotetext{a}{Flux in units of 10$^{-17}$ erg cm$^{-2}$ s$^{-1}$}
\tablenotetext{b}{As defined by Stasinska et al. 2006}
\label{tab:optprop}

\end{deluxetable}

From Table \ref{tab:optprop} it is clear that our k+a selection technique has been effective in isolating galaxies which have undergone recent starbursts. The time since burst of the young k+a sample is $\sim 0.2$ Gyr, the k+a sample $\sim 0.6$ Gyr and the old k+a sample $\sim$ 1--3 Gyr. These values compare well to our predicted time scales of 0.03, 0.3 and 1 Gyr for the 3 classes. Also shown in Table \ref{tab:optprop} is the flux measured in the key AGN diagnostic lines H$\beta$,[OIII],H$\alpha$ and [NII].  By applying AGN diagnostic criteria presented in Stasinska et al. (2006), a variation on the earlier work of Baldwin, Phillips \& Terlevich (1981) \nocite{Stasinska2006, Baldwin1981}, to these we can determine which process dominates the optical emission lines. A clear distinction can be seen between the young k+a galaxies, which are dominated by star formation, and the older classes which are either AGN dominated, or mixtures of AGN and low-level star formation. %However no signs of this AGN activity are observed in the mid-IR spectra, with even the two objects dominated by AGN in their optical spectra (J104953.31+564551.3 \& J104319.20+563151.0) showing no signs of obscured AGN dust tori in the mid-IR. This is confirmed by looking at their position in the IRAC colour-colour diagram of \citet{Lacy2004}, indeed all of the k+a and old k+a galaxies have $\log_{10}$(S$_{5.8}$/S$_{3.6}$)$<-0.2$ placing them firmly within the stellar dominated region. 

\begin{deluxetable}{llllllllll}
\tabletypesize{\tiny}
%\tablewidth{0pt}
\tablecaption{PAH line Fluxes and Equivalent Widths}

\tablehead{
\colhead{ID}  & \colhead{Name}  & \multicolumn{2}{c}{6.2$\mu$m}    &  \multicolumn{2}{c}{7.7$\mu$m} & \multicolumn{2}{c}{8.6$\mu$m}  &  \multicolumn{2}{c}{11.3$\mu$m}   \\
\cline{3-4} \cline{5-6} \cline{7-8} \cline{9-10} 
 & &\colhead{Flux\tablenotemark{a}} & \colhead{EW} & \colhead{Flux} & \colhead{EW} & \colhead{Flux} & \colhead{EW} & \colhead{Flux} & \colhead{EW}\\}

\startdata
\multicolumn{10}{l}{Young k+a}\\
\hline
Yka1 & J103254.44+581638.5  & $1.7\pm0.8$ & 1.99 & $5.1\pm0.8$ & 5.21 & $1.4\pm0.2$ & 1.26 & $1.8\pm0.8$ & 1.02\\
Yka2 & J104826.85+564026.8  & $0.4\pm0.3$ & 0.36 & $1.8\pm0.3$ & 1.11 & $0.5\pm0.09$ & 0.35 & $0.7\pm0.4$ & 0.56\\
Yka3 & J105326.83+580956.5  & $4.7\pm1.0$ & 1.92 & $9.2\pm1.2$ & 3.54 & $2.4\pm0.1$ & 1.0 & $3.7\pm0.8$ & 1.6\\
Yka4 & J163332.66+404827.0  & $2.0\pm0.5$ & 2.92 & $4.2\pm0.6$ & 3.94 & $1.0\pm0.2$ & 0.81 & $1.4\pm0.7$ & 0.65\\

\hline
\multicolumn{10}{l}{k+a}\\
\hline
Ka2 & J104953.31+564551.3  & $0.5\pm0.2$ & 0.45 & $2.1\pm0.3$ & 1.8 & $0.9\pm0.2$&0.75  & $1.1\pm0.3$ & 0.95\\
Ka3 & J110214.89+574630.6  & $3.6\pm0.4$ & 1.2 & $11.0\pm1.4$ & 3.6 & $2.6\pm0.2$ & 0.91 & $4.1\pm0.8$ & 1.3\\
Ka4 & J104319.20+563151.0  & $1.8\pm0.4$ & 0.76 & $5.0\pm0.7$ & 2.5 & $1.3\pm0.2$ & 0.67 & $3.1\pm0.8$ & 1.4\\

\hline
\multicolumn{10}{l}{old k+a}\\
\hline
Oka1 &  J163252.97+402056.1 & -- & -- & -- & -- & -- & -- & -- & -- \\

Oka2 & J104314.82+570941.2  & $0.8\pm0.3$ & 0.28 & $3.6\pm0.3$ & 1.7 & $1.0\pm0.08$ & 0.59 & $2.7\pm0.5$ & 1.8\\
Oka3 & J104650.43+575923.9  & $1.1\pm0.2$ & 0.24 & $5.2\pm0.8$ & 1.5 & $1.0\pm0.2$ & 0.32 & $2.2\pm0.6$ & 0.89\\
Oka4 & J104834.84+583311.3  & $1.1\pm0.3$ & 0.30 & $5.3\pm0.9$ & 2.1 & $0.9\pm0.2$ & 0.37 & $2.8\pm0.5$ & 1.2\\

\enddata
\tablenotetext{a}{$10^{-21}$W~cm$^{-2}$}
\label{tab:irprop}
\end{deluxetable}

\section{Discussion} \label{sec:discussion}
\begin{figure}
\includegraphics[scale=0.8,angle=270]{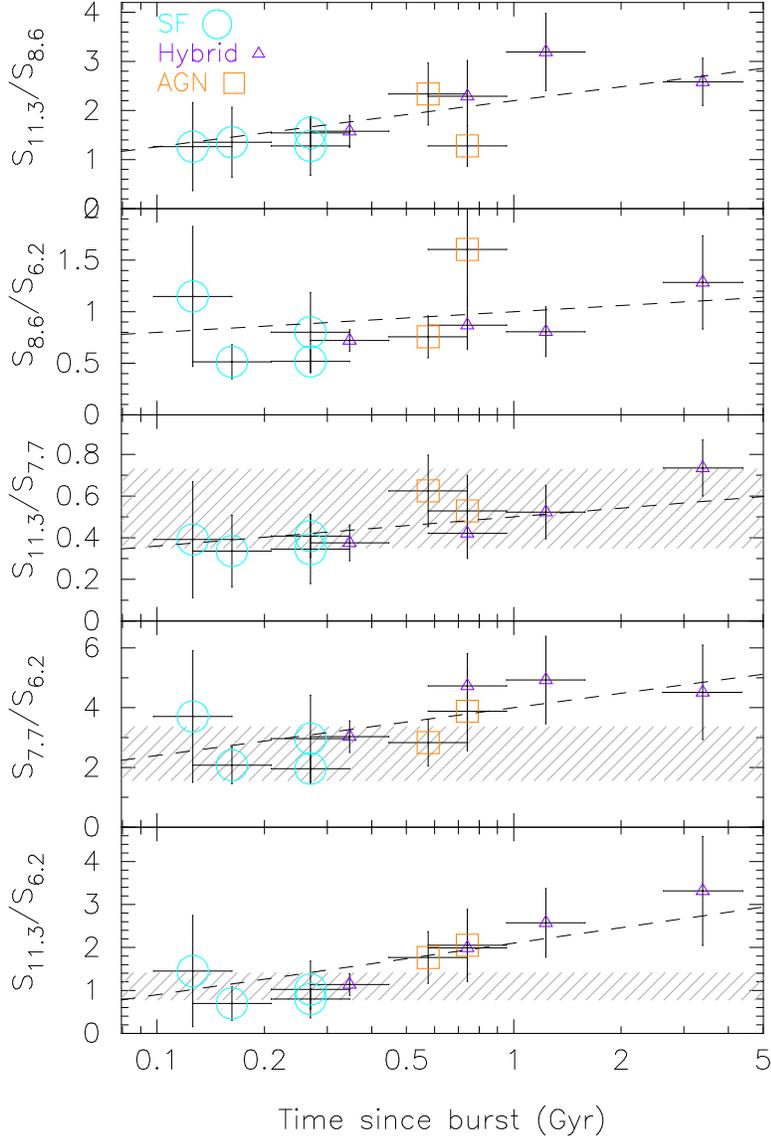}
\caption{Flux ratios of PAH features plotted against the time since last starburst. A trend of increasing S$_{11.3}$/S$_{6.2}$, S$_{7.7}$/S$_{6.2}$ and S$_{11.3}$/S$_{8.6}$ with burst time can be seen. Each point is coded by its optically determined type; circles denote SF dominated systems, squares AGN dominated systems and triangles hybrid combinations of the two. The grey hatched region represents the range of ratios found for starburst galaxies from \citet{Brandl2006}. The dashed line shows the best fit linear relationship between the parameters.}
\label{fig:pahage}
\end{figure}

The unusual nature of the 11.3$\mu$m to 6.2$\mu$m PAH feature in the more evolved k+a galaxies is an interesting discovery. To quantify the nature of this Figure \ref{fig:pahage} shows the ratio of the 6.2$\mu$m flux to the 11.3$\mu$m and 7.7$\mu$m lines as a function of time since starburst.

The S$_{11.3}$/S$_{6.2}$ and S$_{7.7}$/S$_{6.2}$ PAH ratios appear to evolve from the values typical for local starbursts (Brandl et al. 2006) to higher values with increasing time since starburst. Interestingly an equivalent rise is not seen in the S$_{11.3}$/S$_{7.7}$ ratio. %Also the ratios for these objects are well above the 90\% variation limit (1.9) found amongst star forming galaxies in the SINGS survey \citep{Smith2007}. 

To quantify these trends we find the best fit log-linear relationship between the variables. As the errors in both burst time and the flux ratios are significant simple linear regression techniques are not sufficient; hence best fit parameters and their error are determined via a Monte Carlo simulation of the dataset. In each realisation the data points are ``scattered'' by the quoted error multiplied by a normally distributed random number. The analysis here utilises 10,000 realisations. The mean of the slope and intercept across the realisations is then taken to be the best fit value, while the standard deviation is the 1$\sigma$ error on these properties.  Performing this analysis gives the best fit slope (intercept) for the S$_{11.3}$/S$_{6.2}$ ratio vs. $\log_{10}$(Burst Time) of $1.2\pm0.7$ ($2.1\pm0.3$), $1.6\pm1.1$ ($4.0\pm0.5$) for the S$_{7.7}$/S$_{6.2}$, $0.14\pm0.14$ ($0.5\pm0.06$) for the S$_{11.3}$/S$_{7.7}$, $0.2\pm0.3$ ($1.0\pm0.1$) for the S$_{8.6}$/S$_{6.2}$ and $0.94\pm0.58$ ($2.2\pm0.3$) for the S$_{11.3}$/S$_{8.6}$. This suggests that while the observed S$_{11.3}$/S$_{7.7}$ is consistent with no evolution, the S$_{11.3}$/S$_{6.2}$, S$_{7.7}$/S$_{6.2}$ and S$_{11.3}$/S$_{8.6}$ ratios show a statistically significant non-zero evolution with burst time.

Smith et al. (2007) hypothesise that variations in PAH ratios could be generated via either weaker radiation fields, resulting in more neutral PAHs, or preferential destruction of the smallest grains in the presence of a very hard radiation field. They find that the 10 objects in their sample with the highest S$_{11.3}$/S$_{7.7}$ PAH ratios have clear evidence for LINER or Seyfert AGN activity at longer wavelengths, suggesting that hard radiation fields associated with AGN could be preferentially destroying the smaller grains responsible for the 6.2$\mu$m and 7.7$\mu$m bands, leaving an excess of the larger 11.3$\mu$m emitting grains.

Here we also find that all of the galaxies which have high S$_{11.3}$/S$_{6.2}\mu$m PAH ratios have evidence of AGN activity in their optical spectra. Interestingly we do not find that the S$_{11.3}$/S$_{7.7}$ PAH ratio varies with time since starburst in the same way as the S$_{11.3}$/S$_{6.2}$ ratio. 

However this is further complicated by the observed evolution in the S$_{11.3}$/S$_{8.6}$ ratio, which is what would be expected from grain destruction, but should not be possible without an accompanying S$_{11.3}$/S$_{7.7}$ ratio change. Even more challenging is the odd behaviour of the S$_{8.6}$/S$_{6.2}$ ratio, which shows no evolution despite the S$_{7.7}$/S$_{6.2}$ showing a clear change with time since starburst. We put these irregularities down to underestimates in the error in the 8.6$\mu$m line flux, which is heavily blended with the larger 7.7$\mu$m feature and is thus badly effected by the poor estimates of the continuum; a serious possibility given that we only have 5--12$\mu$m spectroscopy for these objects.

However AGN are only one possible explanation for this behaviour of the PAH band ratios. Detailed modelling of the PAH emission models by \citet{Galliano2008} has shown that increasing the hardness of the interstellar radiation field (ISRF) will result in a sharp decrease in both the S$_{11.3}$/S$_{7.7}$ ratio and the S$_{7.7}$/S$_{6.2}$ ratio while destroying the smallest grains ($<10^3$\AA) will have the inverse affect. In addition increasing the ionization level of the PAH emitting grains will decrease the S$_{11.3}$/S$_{7.7}$ ratio while leaving the S$_{7.7}$/S$_{6.2}$ ratio unaffected.

Thus the results found here can be explained via the following;
\begin{enumerate}
\item The ionization level is higher than normal in these objects, decreasing the S$_{11.3}$/S$_{7.7}$ ratio
\item However the ionization level is so high that the smallest grains are being destroyed, increasing both the S$_{11.3}$/S$_{ 7.7}$ \& S$_{7.7}$/S$_{6.2}$ (and thus the S$_{11.3}$/S$_{ 6.2}$ as well).
\end{enumerate}
Thus the S$_{11.3}$/S$_{ 7.7}$ ratio may be balanced out by the competing processes and appear to have ``normal'' values, while the S$_{11.3}$/S$_{6.2}$ \&  S$_{7.7}$/S$_{ 6.2}$ ratios show abnormal increases. While any source which can strongly ionize and destroy the PAH emitting grains could be invoked to explain this phenomena, given the evidence of low-level AGN activity in the optical spectra for these sources we conclude that this is the most likely cause for these unusual PAH line ratios. However this is purely speculative and other ionization sources such as significant populations of evolved stars may be capable of the levels of grain destruction and ionization seen here. The behaviour of the 8.6$\mu$m line ratios does not fit into this model, as we should see similar behaviour from the S$_{8.6}$/S$_{6.2}$ ratio as the S$_{7.7}$/S$_{6.2}$, which we do not. 

\section{Conclusion}
We have performed {\it Spitzer} IRS spectroscopy of 10 post starburst (k+a) galaxies at three distinct evolutionary stages. In each case the mid-IR spectra shows the signs of star forming galaxy with no obvious AGN activity. Interestingly we find that relative strength of the 11.3$\mu$m to 6.2$\mu$m, 11.3$\mu$m to 8.6$\mu$m and 7.7$\mu$m to 6.2$\mu$m PAH features show some correlation with the time since last starburst, while the 11.3$\mu$m to 7.7$\mu$m  and  S$_{8.6}$/S$_{6.2}$ PAH ratios do not. We conclude that this behaviour is most likely a result of underlying low-level AGN activity in these objects.
\acknowledgments
This work is based on observations made with the Spitzer Space Telescope, which is operated by the Jet Propulsion Laboratory, California Institute of Technology under a contract with NASA. \\

The IRS was a collaborative venture between Cornell University and Ball Aerospace Corporation funded by NASA through the Jet Propulsion Laboratory and Ames Research Center.\\

SMART was developed by the IRS Team at Cornell University and is available through the Spitzer Science Center at Caltech. \\
We thank C. Maraston for useful conversations regarding the use of stellar population synthesis models which greatly aided the paper. We also thank M. Polletta for useful contributions to this work.\\
We thank the anonymous referee for comments which greatly improved this work.\\

This work was supported by the Science and Technology Facilities  Research Council [grant number ST/F002858/1].\\

{\it Facilities:} \facility{Spitzer(IRS)}, \facility{SDSS}.

%\bibliography{lit}
\bibliographystyle{apj}

\end{document}